\documentclass[12pt,a4paper,english]{article}
\usepackage[T1]{fontenc}
\usepackage[latin1]{inputenc}

\makeatletter

\providecommand{\LyX}{L\kern-.1667em\lower.25em\hbox{Y}\kern-.125emX\@}
\usepackage{graphicx,epsfig}
\usepackage{color}

\newcommand{\gsim}{\, \mathop{}_{\textstyle \sim}^{\textstyle >} \,}
\newcommand{\lsim}{\, \mathop{}_{\textstyle \sim}^{\textstyle <} \,}

\usepackage{babel}
\makeatother
\begin{document}

\title{\textbf{\large Dark Matter from an ultra-light pseudo-Goldsone-boson}\footnote{CERN-PH-TH/2005-163}}

\author{\textit{\emph{Luca Amendola$^{a}$ and Riccardo Barbieri$^{b,c}$}}\\
 $^{a}$\emph{INAF/Osservatorio Astronomico di Roma, V. Frascati 33,
00040} \\
 \emph{Monteporzio Catone (Roma), Italy}\\
 \textit{$^{b}$} \textit{Scuola Normale Superiore and INFN, Piazza
dei Cavalieri 7,} \\
 \textit{I-56126 Pisa, Italy}\\
 \emph{${}^{c}$Theoretical Physics Division, CERN, CH-1211} \\
 \emph{Gen\`{e}ve 23, Switzerland}\\
 }

\maketitle
\begin{abstract}
Dark Matter (DM) and Dark Energy (DE) can be both described in terms
of ultra-light Pseudo-Goldstone-Bosons (PGB) with masses $m_{DM}\gsim 10^{-23}$eV
and $m_{DE}\lsim 10^{-33}$eV respectively. Following Barbieri et
al, we entertain the possibility that a PGB exists with mass $m_{I}$
intermediate between these two limits, giving a partial contribution
to DM. We evaluate the related effects on the power spectrum of the
matter density perturbations and on the cosmic microwave background
and we derive the bounds on the density fraction, $f_{I}$ , of this
intermediate field from current data, with room for a better sensitivity
on $f_{I}$ in the near future. We also give a simple and unified
analytic description of the free streaming effects both for an ultra-light
scalar and for a massive neutrino. 
\end{abstract}

\section{Introduction and statement of the problem}

To understand the nature of Dark Matter (DM) and Dark Energy (DE)
is one of the most intriguing problems in all of today's physics.
Proposals for possible solutions are not lacking. More difficult is
to prove any of them by a direct signal; an example in the case of
DM would be the detection in the laboratory of a WIMP interaction.
In this note we pursue the idea that part of DM may be due to an ultra-light
Pseudo-Goldstone-Boson (PGB), arising from the spontaneous breaking
at a scale close to $M_{Pl}$ of an extended approximate symmetry,
and we study a possible related signal.

The idea that DM or DE may be interpreted as the energy density of
a PGB is not new. The axion is the prototype example for DM. Well
known is also the fact that the mass of the hypothetical scalar field
associated to DE would have to satisfy the bound $m_{DE}\lsim 10^{-33}eV$.
The tightness of this bound provides in fact a clear case for interpreting
also this scalar as a PGB\cite{Weiss:1987xa,Frieman:1995pm,PGBDE}.

Following this view, a calculable microscopic model for the potential
of a PGB for DE has been proposed in Ref. \cite{Barbieri:2005gj} , based
on an approximate $U(1)^{n}$ flavour symmetry of the right-handed-neutrinos.
For the purposes of the present paper, the key feature of the model
is the presence of several PGBs, $G_{i}$, up to five in the general
version, with a potential of the form \begin{equation}
V(G_{i})=\Sigma _{i}\mu _{i}^{4}(\cos {(G_{i}/F_{i})}+1),\label{eq:VPGB}\end{equation}
 where $F_{i}$ are mass scales related to the scales of spontaneous
breaking of the $U(1)$ factors and the heights of the various potential
terms, $\mu _{i}^{4}$, are controlled by small parameters that explicitly
break the $U(1)^{n}$ flavour symmetry. For the interpretation of
DE, one such term, with $i\equiv DE$, must have $G_{DE}^{in}\approx F_{DE}\approx M_{Pl}$
and $\mu _{DE}\approx 3\cdot 10^{-3}eV$, with $G_{DE}^{in}$ the
initial value of the DE-PGB field. On the other hand, the presence
of other similar terms in eq. (\ref{eq:VPGB}), with $i\neq DE$,
makes it natural to ask what the manifestation of the other PGBs could
be. If their masses, $m_{i}\approx \mu _{i}^{2}/F_{i}$, are greater
than $m_{DE}\approx H_{0}$, the Hubble constant today, then these
extra PGBs contribute to DM, with possible effects on the growth of
cosmic clustering. One such scalar, of mass $m_{DM}\approx 10^{-23}\div 10^{-22}eV$,
has in fact already been invoked with the purpose of suppressing the
apparently unobserved cusps arising in the halos of conventional Cold
Dark Matter (CDM) \cite{Hu:2000ke}. In the potential of eq.~(\ref{eq:VPGB}),
the parameters that would lead to the interpretation of DM in terms
of such a field are $G_{DM}^{in}\approx F_{DM}\approx 0.01M_{Pl}$
and $\mu _{DM}\approx 30$ eV.

Here we entertain the possibility that a PGB exists with a mass $m_{I}$
intermediate between $m_{DM}\sim 10^{-23}eV$ and $m_{DE}\sim 10^{-33}eV$.
In the range of parameters \begin{equation}
\frac{F_{I}}{M_{Pl}}=0.01\div 1;\quad \mu _{I}=3\cdot 10^{-3}\div 30eV\end{equation}
 one would have indeed \begin{equation}
m_{I}=10^{-33}\div 10^{-23}eV;\quad f_{I}\equiv \frac{\Omega _{I}}{\Omega _{m}}\sim 10^{-3}\div 1\end{equation}
 where $f_{I}$ is the fraction of the energy density in the field
$G_{I}$ relative to the matter density, possibly given by a heavier
PGB. In general, for $G_{I}^{in}\approx F_{I}$, it is, when $f_{I}<1$,
\begin{equation}
f_{I}\approx \frac{8\pi }{3}(\frac{F_{I}}{M_{Pl}})^{2}max(1,(\frac{m_{I}}{H_{eq}})^{1/2})\label{f_I}\end{equation}

The interest of the presence of such a field is that it would lead
to peculiar features of the matter power spectrum, quite analogous
to but also generally different from the ones of a massive neutrino.
As we now recall, these effects are related to the wave properties
of the fluid associated with an ultra-light scalar, which inhibit
the clustering of its energy density at sufficiently small scales.
The detection of these effects might therefore constitute a signal
of the overall underlying picture.

The CMB is also affected by such a field. In fact, when the expansion
rate $H$ is higher than $m_{I}$, the field freezes and behaves as
a cosmological constant. If this occurs at the scale factor $a_{F}$ before the present,
the field behaves for $a<a_{F}$ as a cosmological constant with density
fraction $\Omega _{I<}(a)\approx (a/a_{F})^{3} f_I \Omega _{m}$
(assuming matter dominates). If $\Omega _{I<}(a)$ is not negligible at
decoupling time, then there is a distortion in the anisotropy spectrum
due to the early Sachs-Wolfe effect. Since decoupling occurs at $H_{d}\approx 6\cdot 10^{-29}$eV, 
we expect the CMB distortion will be maximal for $m_{I}$ near this
value.

\section{Free streaming in an analytic approximation}

The physics underlying these phenomena is well known\cite{Zeldovich,Ferreira:1997hj,Hu:2000ke}.
A straightforward modification of the CMBFAST software \cite{sz}
indeed allows a numerical calculation of these effects (See Sect.
3). To ease the reading and to make explicit their connection with
the effects of a massive neutrino, we derive them in a simple and
effective analytic approximation.

What makes the difference between an ultra-light scalar or a massive
neutrino and ordinary CDM is the presence of a pressure term in the
Euler equation for the equivalent fluids, $(k^{2}/a^{2})c_{eff}^{2}\delta _{k}$,
where $\delta _{k}$ is the density perturbation of the mode with
comoving wave number $k$, $a$ is the scale factor and $c_{eff}$
is an effective sound speed, which becomes less then 1 at sufficiently
large $a$. For a PGB of mass $m_{\phi }$ it is 
\begin{equation}
c_{eff}\approx \frac{k}{2am_{\phi }}\quad \textrm{for}\quad a\gsim \frac{k}{2m_{\phi }},
\label{eq:scalar}
\end{equation}
 whereas for a neutrino of mass $m_{\nu }\cite {Bond:1983hb,Ma:1996za}$
 \begin{equation}
c_{eff}\approx \frac{T_{\nu }^{0}}{m_{\nu }a}\quad \textrm{for}\quad a\gsim \frac{T_{\nu }^{0}}{m_{\nu }}\equiv a_{NR}
\label{eq:nu}
\end{equation}
 where $T_{\nu }^{0}$ is the neutrino temperature today. This pressure
term introduces a Jeans length, $1/k_{J}$, below which the fluids are
unable to cluster: they \char`\"{}free stream\char`\"{}. Equating
the pressure term to the source term, $4\pi G\rho \delta _{k}$, during
matter domination, one finds \begin{equation}
k_{J}^{\phi }(a)=1.56a^{1/4}m_{\phi }^{1/2}H_{0}^{1/2}\Omega _{m}^{1/4}\quad \textrm{and}\quad k_{J}^{\nu }(a)=1.22a^{1/2}\frac{m_{\nu }}{T_{\nu }^{0}}H_{0}\Omega _{m}^{1/2}.\label{eq:nu}\end{equation}

A classic result is that, if all matter contributing to the cosmic
density is able to cluster, like ordinary CDM, the corresponding density
fluctuations, $\delta _{k}^{c}$, all grow like the scale factor,
$\delta _{k}^{c}\propto a$, between matter-radiation equality, $a=a_{eq}$,
and now, $a=a_{0}$. Similarly, if only a fraction $(1-f)$ can cluster,
the growth is slower\cite{Bond:1980ha}\begin{equation}
\delta _{k}^{c}\propto a^{p},p=1/4(\sqrt{1+24(1-f)}-1).\label{eq:growth}\end{equation}
 This means that, relative to the case in which free streaming is
neglected, the power spectrum of the matter density fluctuations is
given by 
\begin{equation}
r(k)\equiv \frac{P_{true}}{P_{no-free-streaming}}=(\frac{a_{max}}{a_{min}})^{2(p-1)}
\label{eq:P_rel}
\end{equation}
 where $(a_{min},a_{max})$ is the $k$-dependent interval of scale
factor during which free streaming is relevant.

In the case of the PGB, defining $a_{m}$ as the scale factor at which
the field starts to oscillate, at $H\approx m_{\phi }$, and $a_{J}^{\phi }(k)$
as the scale factor at which $k=k_{J}^{\phi }(a)$, it is $a_{min}=max(a_{m},a_{eq})$
and $a_{max}=min(a_{0},a_{J}^{\phi }(k))$. The relevant scales are
therefore $(m_{30}\equiv m_{\phi }/(10^{-30}eV)$) \begin{equation}
k_{eq}^{\phi }\equiv k_{J}^{\phi }(a_{eq})\approx 9\cdot 10^{-4}m_{30}^{1/2}Mpc^{-1}\label{eq:nu}\end{equation}
\begin{equation}
k_{0}^{\phi }\equiv k_{J}^{\phi }(a_{0})\approx 1.1\cdot 10^{-2}m_{30}^{1/2}(h^{2}\Omega _{m})^{1/4}Mpc^{-1}\label{eq:nu}\end{equation}
\begin{equation}
k_{m}^{\phi }\equiv k_{J}^{\phi }(a_{m})\approx 3.6\cdot 10^{-3}m_{30}^{1/3}(h^{2}\Omega _{m})^{1/3}Mpc^{-1}\label{eq:nu}\end{equation}
 and, $r^{\phi }$, unaffected below $k_{min}^{\phi }$, for $k>k_{min}^{\phi }$
becomes \begin{equation}
r^{\phi }(k)\approx (\frac{k_{min}^{\phi }}{k_{max}^{\phi }})^{8(1-p)}\label{eq:P_phi}\end{equation}
 where $k_{min}^{\phi }=max(k_{eq}^{\phi },k_{m}^{\phi })$ and $k_{max}^{\phi }=min(k,k_{0}^{\phi })$.

In the neutrino case, one proceeds in a fully analogous way. Let us
restrict ourselves to the case $m_{\nu }<1eV$, so that the neutrino
becomes non relativistic after matter-radiation equilibrium, $a_{NR}>a_{eq}$.
In this case there are only two relevant scales ($m_{eV}\equiv m_{\nu }/eV$)
\begin{equation}
k_{NR}^{\nu }\equiv k_{J}^{\nu }(a_{NR})\approx 0.026m_{eV}^{1/2}(h^{2}\Omega _{m})^{1/2}Mpc^{-1}\label{eq:nu}\end{equation}
\begin{equation}
k_{0}^{\nu }\equiv k_{J}^{\nu }(a_{0})\approx 2.2m_{eV}(h^{2}\Omega _{m})^{1/2}Mpc^{-1}\label{eq:nu}\end{equation}
 and, for $k>k_{NR}$, \begin{equation}
r^{\nu }(k)\approx (\frac{k_{NR}}{k_{max}^{\nu }})^{4(1-p)}\label{eq:P_nu}\end{equation}
 where $k_{max}^{\nu }=min(k,k_{0}^{\nu })$. 

Eq.s~(\ref{eq:P_phi}, \ref{eq:P_nu}) are the reference formulae.
They are immediately useful when the power spectrum with neglect of
free streaming is simple to compute, since only then, from eq.~(\ref{eq:P_rel}),
$P_{true}$ is easily determined from $r(k)$. This is so if the fraction
$f$, relative to conventional CDM, of the extra component we are
considering is small, below $10\%$, which is anyhow the case of interest
here. If so, $r^{\phi }(k)$ in eq.~(\ref{eq:P_phi}) and $r^{\nu }(k)$
in eq.~(\ref{eq:P_nu}) give directly the power spectrum normalized
to conventional $\Lambda $CDM with 3 massless neutrinos. In principle,
this is not precise. In the low mass range, at sufficiently early
times but still after equilibrium, neither the PGB nor the neutrinos
behave as CDM, even apart from free streaming. Since massless neutrinos
are anyhow included, this is a negligible effect for neutrinos. In
the PGB case, a better approximation of the denominator in eq. (\ref{eq:P_rel}),
especially at small scales, consists indeed in including the contribution
of a fictitious scalar field with $c_{eff}=0$.

Note that in the neutrino case, its mass determines as well $f$,
which appears in the exponent of eq.~(\ref{eq:growth}), according
to 
\begin{equation}
f\equiv \frac{\Omega _{\nu }}{\Omega _{m}}=\frac{0.011}{h^{2}\Omega _{m}}m_{eV} \equiv f_\nu.
\label{eq:f_nu}
\end{equation}
 Note also that, if more than one neutrino is massive, all below 1
eV so that each relative $f$ is small, one obtains the power spectrum
normalized to the massless case by taking a factor as in eq.~(\ref{eq:P_nu})
for every massive neutrino.

Figs. 1 and 2 show the power spectrum of density perturbations for
the PGB (Fig. 1) and the massive neutrino cases (Fig. 2), normalized
as described above, for some representative values of the relevant
parameters. The continuous lines are from the numerical code, whereas
the dashed lines represent the analytic expressions discussed above.
The analogies and the differences between the PGB and the neutrino
case are apparent from the figures themselves and from the discussion
above. Neutrino masses are known to exist at a level that might be
within reach of future cosmological measurements. In the ideal case
in which all the various parameters of $\Lambda $CDM were known,
there are two characteristic differences between Fig. 1 and 2, that
can be used to distinguish them. The interval in $k$ with a non-vanishing
slope is controlled by the mass parameters in a characteristically
different way for the neutrino and the PGB cases. The slope itself
depends on the fraction $f$, which in the PGB case is a free parameter
whereas for the neutrinos it is tied to the mass by eq. (\ref{eq:f_nu}).

\section{Comparison with observations}

We now compare with current observations the matter power spectrum
and CMB when the extra PGB scalar is present along with a cosmological
constant and ordinary CDM. Both the cosmological constant and ordinary
CDM could be effective descriptions of ultra-light PGBs, of appropriate
mass, but this is irrelevant to the discussion of this Section. As
recalled in the previous Section, the perturbation equations for a
generic scalar field can be expressed as a perfect fluid with a scale-dependent
sound speed. Moreover, we take the field behaving as a cosmological
constant ($w_{I}=-1$) for $H>m_{I}$ and as dark matter ($w_{I}=0$)
afterward.

We produce the power spectra and the temperature and polarization
CMB spectra by using a suitably modified version of CMBFAST \cite{sz}
which includes the extra fluid. We explore the range of masses and
densities\begin{eqnarray*}
m_{I} & = & (10^{-6}\div 10^{8})m_{30}\\
f_{I} & = & 0\div 1
\end{eqnarray*}
 which includes as extreme cases the behavior of a cosmological constant
(masses smaller than $H_{0}\approx 10^{-33}$eV ) and of ordinary
cold dark matter (masses larger than the typical galaxy size, $10^{-24}$eV).

We compare the results via a grid-based likelihood analysis with four
observations: the power spectrum of SDSS \cite{sdss}( cutting at
$k=0.2$Mpc/$h$); the power spectrum of Lyman-$\alpha $ clouds \cite{croft};
the variance $\sigma _{8}=0.90\pm 0.03$ derived by \cite{selj} combining
Lyman-$\alpha $ data in the SDSS catalog with WMAP data. Finally,
we use the CMB temperature and polarization spectra of the 2001 WMAP
dataset \cite{wmap}. Notice that the datasets are all independent
except for the use of WMAP to fix the initial perturbation amplitude
in deriving $\sigma _{8}$: when we combine all constraints we neglect
this residual correlation. (It is also to be noted that $\sigma _{8}$
is subject to considerable systematic uncertainty \cite{viel}, potentially
quite larger than the statistical one.) For the SDSS we used the likelihood
routine provided by M. Tegmark which includes the full covariance
matrix while for CMB data we adopt the likelihood routine provided
by the WMAP team \cite{verde}. When comparing to SDSS and Lyman-$\alpha $
spectra we marginalize over the amplitudes (separately for the two
datasets), so that the information on the spectrum amplitude is used
only in the third test. In this way the large-scale structure constraints
are well separated into those derived from the large scale slope (SDSS),
the small scale slope (Lyman-$\alpha $) and the amplitude ($\sigma _{8}$).
Finally, we combine all four tests to give an exclusion plot for $m_{I},f_{I}$.

We let the other cosmological parameters vary in the ranges \begin{eqnarray*}
h & \in  & 0.6\div 0.8\\
\Omega _{m}h^{2} & \in  & 0.06\div 0.15\\
\tau  & \in  & 0\div 0.3\\
n_{s} & \in  & 0.95\div 1.05
\end{eqnarray*}
 while to save computer time we fix  the baryon fraction $\Omega _{b}h^{2}=0.023$.
We assume flat priors for all parameters.

The results are shown in Figs. 3 (the likelihood for each dataset)
and 4 (the combined likelihood), after marginalization over the other
cosmological parameters (and the overall normalization for WMAP, SDSS
and Lyman-$\alpha $). The Lyman-$\alpha$ constraints turn out to be the strongest. As expected, there is an intermediate-mass
window of observability between $10^{-31}$ eV and $10^{-23}$eV. In this region of interest for the mass $m_{I}$,
the relative density $f_{I}$ is clearly limited to be below $10\%$,
which is a significant constraint but leaves ample room for a relevant
component of a DM-PGB, as given in eq. (\ref{f_I}).

The prospects for detecting the signal from an intermediate-mass PGB
scalar are interesting. The same cosmological observations that will
constrain the neutrino mass can provide limits to the PGB mass and
abundance, since the PGB scalar affects both the growth of fluctuations
and the power spectrum shape. In particular, deep weak lensing observations
are expected to provide in few years upper limits to the neutrino
abundance $f_{\nu }$ of  less than one percent \cite{dod}, mostly
through the damping effect on the fluctuation growth. Since a scalar
with the same abundance $f_{I}$ provides a very similar fluctuation
growth (at least for $m_{I}\in 10^{-30}\div 10^{-24}$eV) we can expect
that also $f_{I}$ can be constrained to better than the percent accuracy.
This similarity of behavior raises in fact also the issue of the level of degeneracy
of scalars with neutrinos, a problem which may deserve
attention.

\begin{figure}
\includegraphics[ bb=0bp 200bp 612bp 792bp,
  clip, scale=0.6]{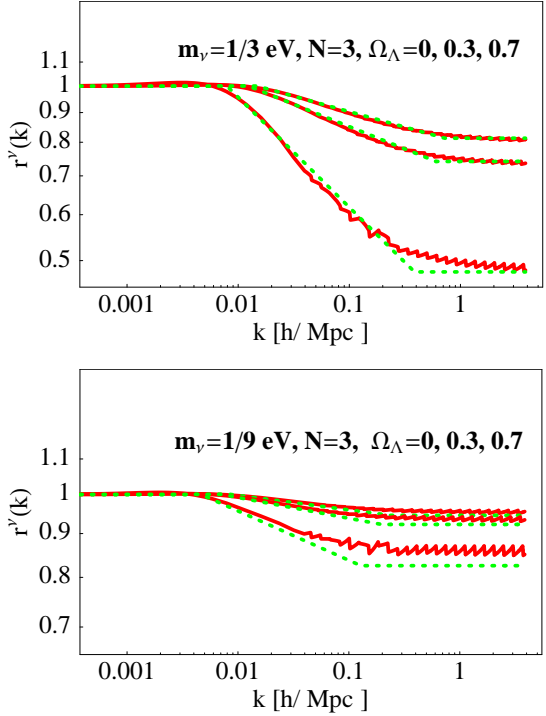}

\caption{Comparison of the ratios $r^{\nu }(k)$ of numerical matter power
spectra (red full curves) with the analytic approximation discussed
in the text (green dashed curves). Upper panel: 3 massive neutrinos
with mass 1/3 eV each, and $\Omega _{\Lambda }=0,0.3,0.7$, (ordered
top to bottom according to the left plateau). Lower panel: same, with
masses 1/9 eV.}
\end{figure}

\begin{figure}
\includegraphics[  bb=0bp 200bp 612bp 792bp,
  clip,
  scale=0.6]{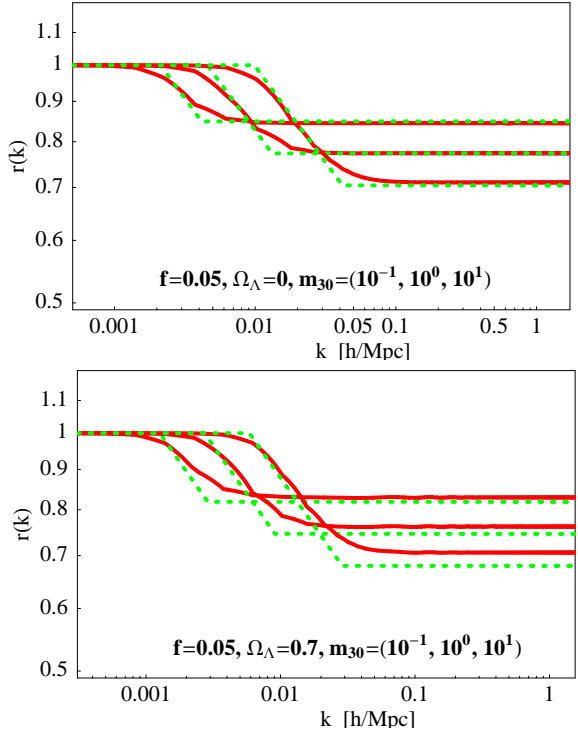}

\caption{Comparison of the ratios $r^{\phi }(k)$ of numerical matter
power spectra (red full curves) with the analytic approximation discussed
in the text (green dashed curves). Upper panel: PGB with fraction
$f_{I}=0.05$ and masses $m_{30}=0.1,1,10$ (ordered top to bottom
according to the left plateau), and $\Omega _{\Lambda }=0$. Lower
panel: same, with $\Omega _{\Lambda }=0.7$.}
\end{figure}

\begin{figure}
\includegraphics[  bb=0bp 200bp 612bp 792bp,
  clip,
  scale=0.8]{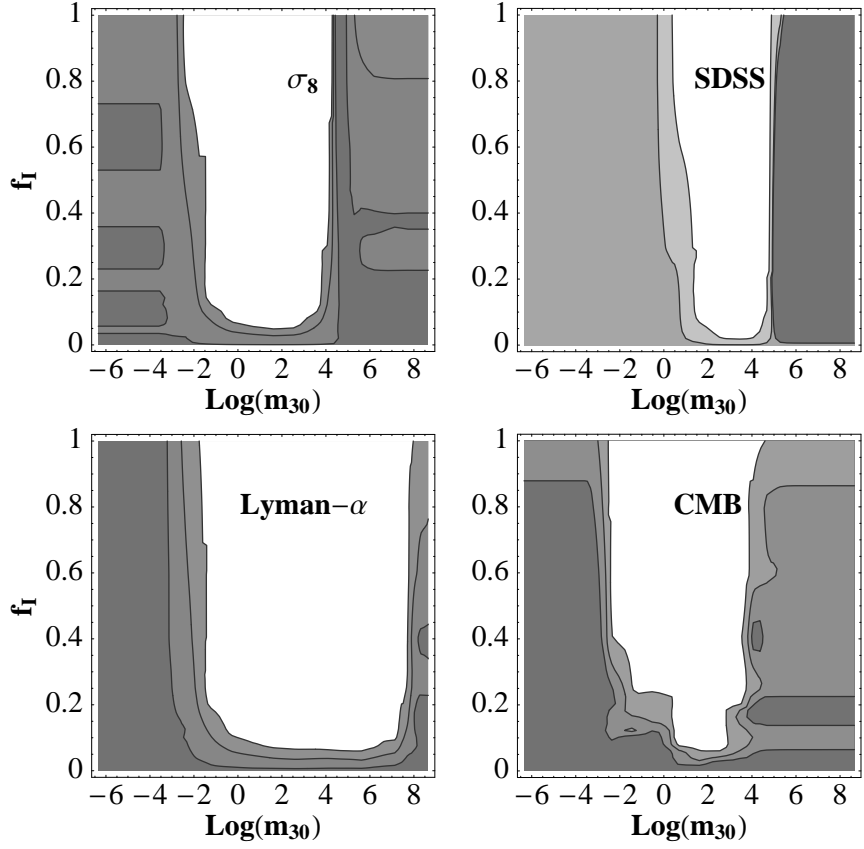}

\caption{Likelihood functions at 68,95 and 99.7\% c.l. (dark to light gray)
for the parameters $m_{30}\equiv m_{I}/10^{-30}eV$ and $f_{I}$ obtained
marginalizing over $\tau ,\Omega _{m}h^{2},n_{s}$ and $h$, and fixing
$\Omega _{b}h^{2} = 0.023$. The date employed are discussed in the text.
The discreteness of the grid causes some wiggling in the function.}
\end{figure}

\begin{figure}
\includegraphics[  bb=0bp 200bp 612bp 792bp,
  clip,
  scale=0.6]{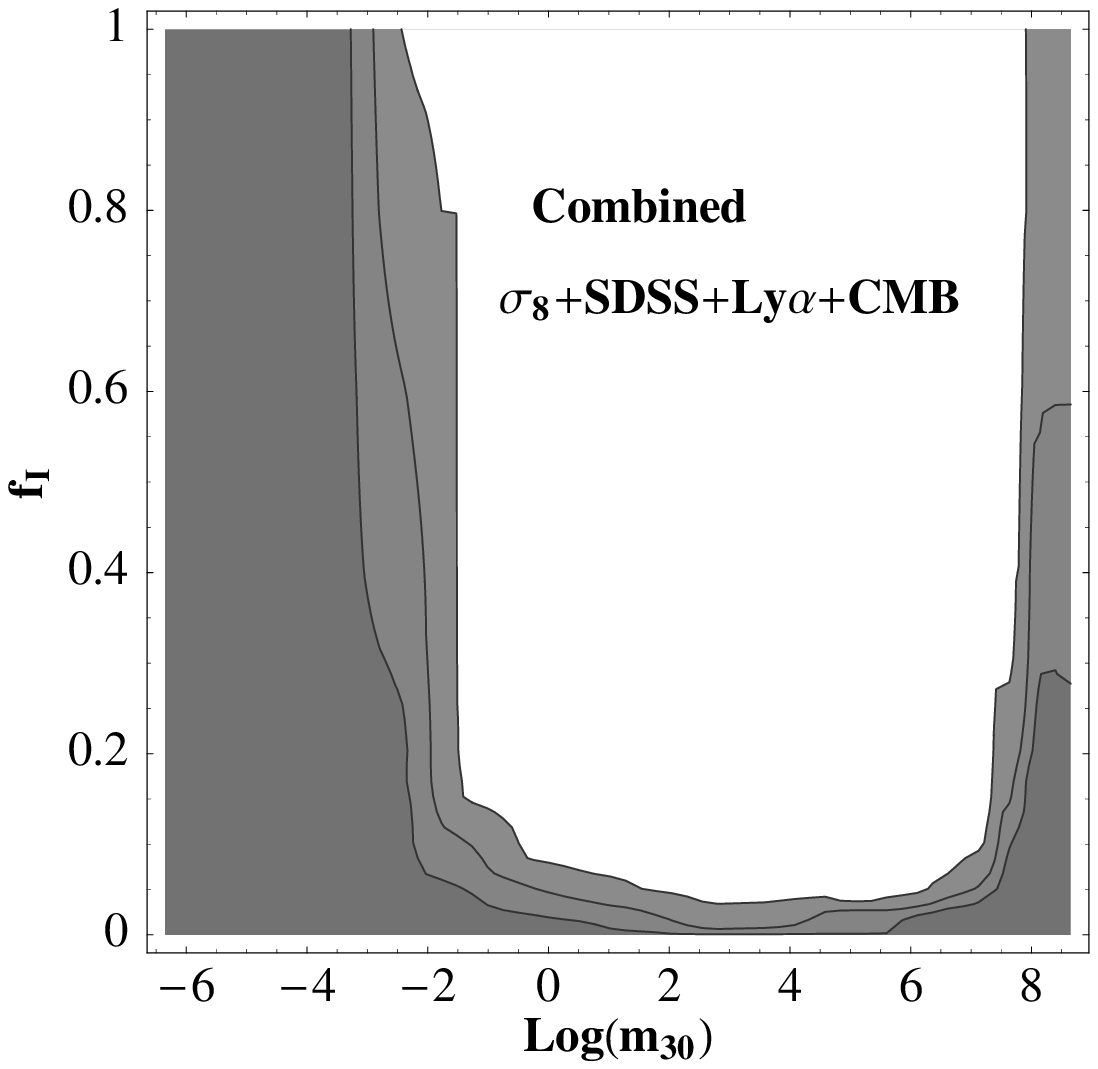}

\caption{Combined likelihood function at 68,95 and 99.7\% c.l. (dark to light
gray)  for the parameters $m_{30}\equiv m_{I}/10^{-30}eV$ and $f_{I}$
obtained marginalizing over $\tau ,\Omega _{m}h^{2},n_{s}$ and $h$,
and fixing $\Omega _{b}h^{2}= 0.023$. }
\end{figure}

\section*{Acknowledgments}

This work is supported in part by MIUR and by the EU under RTN contract
MRTN-CT-2004-503369.

\end{document}